\documentclass[aps,prd,twocolumn,superscriptaddress,showpacs]{revtex4-2}
\usepackage{color}
\usepackage{latexsym}
\usepackage{amsmath}
\usepackage{amssymb}
\usepackage{eufrak}
\usepackage{euscript}
\usepackage[latin1]{inputenc}
\usepackage{pstricks}
\usepackage{graphics}
\usepackage{graphicx}
\usepackage{picture}
\usepackage{pstricks,pst-coil}
\usepackage{pst-all}
\usepackage{amsmath,amssymb,mathtools}
\usepackage[margin=1in]{geometry}
\usepackage{bm}

\usepackage{graphicx} 

\begin{document}

\title{One-loop effective potential in Kalb-Ramond scalar electrodynamics}
%
%

\author{Pratik Chattopadhyay}\email{pratikpc@gmail.com}

\affiliation{Physics Department, Ariel University, Ariel 40070, Israel}

\author{M. J. Neves}\email{mariojr@ufrrj.br}

\affiliation{Departamento de F\'isica, Universidade Federal Rural do Rio de Janeiro, BR 465-07, 23890-971, Serop\'edica, Rio de Janeiro, Brazil}
\begin{abstract}
In this work, we study the effective potential at one loop for the Kalb-Ramond scalar electrodynamics. The model is based on a complex scalar field coupled to the electromagnetic field as usual, and also to the Kalb-Ramond field through the dual vector associated with the strength tensor of the $2$-form gauge field. There is an additional topological coupling of the electromagnetic field and the Kalb-Ramond field. The quantum corrections generated by the Kalb-Ramond sector are computed using dimensional regularization, and the ultraviolet divergences are removed by introducing the appropriate counterterms, leading to a finite renormalized effective potential. We find that the effective potential at one-loop generates terms of the form $\phi^6$, which is not present in the original bare Lagrangian. As a result, we introduced new counter terms to eliminate such divergences. The theory thus considered is not closed under renormalization.

\end{abstract}
\maketitle

\section{Introduction}
The Kalb-Ramond is a two-form gauge field that emerges naturally as a massless excitation of the closed string, in addition to the graviton and dilaton \cite{Kalb}. Unlike point particles, which couple naturally to a vector gauge field through their one-dimensional worldlines, strings sweep out two-dimensional worldsheets, and therefore they interact minimally with an antisymmetric rank-two gauge potential, namely the Kalb-Ramond field \cite{GSW,Banks}. The latter also appears in the low-energy limit of superstring theories as the Neveu-Schwarz two-form of supergravity, and has subsequently found applications in many research areas like higher-form gauge theories, topological field theory, cosmology, black hole dynamics and effective descriptions beyond the Standard Model, which includes spontaneous symmetry breaking
\cite{Dass,Esposito,Capanelli,Turaev,Barone,Rey}. The dimensional reduction of the three-form model from five to four dimensions also opens the investigation of exotic fields coupled to Kalb-Ramond candidates to a dark sector \cite{Neves}. In the functional formalism of quantum field theory (QFT), the effective potential is constructed from the effective action which is one of the tools for investigation of ultraviolet divergences and renormalization \cite{CW,Chattopadhyay}.

Motivated by these developments, it is natural to investigate the quantum effects induced by a Kalb-Ramond sector when coupled to charged scalar matter. While previous studies have largely focused on classical dynamics and topological mass generation, very little attention has been devoted to their influence on the quantum vacuum. Since the effective potential encodes the radiative corrections to the vacuum energy and determines the pattern of spontaneous symmetry breaking, it provides the appropriate framework for analyzing these effects. In this work, we compute and renormalize the one-loop effective potential of scalar electrodynamics supplemented by a gauge-invariant coupling to the dual Kalb-Ramond field strength, and investigate the resulting modifications of the vacuum structure induced by the higher-form gauge interaction.

%
%

%
The work is organized as follows : In section (\ref{sec2}), we present the Kalb-Ramond-Maxwell coupled to a complex scalar field. In section (\ref{sec3}) we apply the functional formalism to obtain the integrals for the effective potential. Section (\ref{sec4}) uses the dimensional regularization to isolate divergent terms. In section (\ref{sec5}), we renormalize the effective potential. In the end, the conclusions are presented in section (\ref{sec6}).  
We use the natural units of $c=\hbar=1$, with the Minkowski metric of signature $\eta^{\mu\nu}=(+1,-1,-1,-1)$ throughout this paper. 

\section{The Kalb-Ramond scalar electrodynamics}
\label{sec2}

The Kalb-Ramond scalar electrodynamics is set by the Lagrangian density
\begin{eqnarray}\label{LKB}
{\cal L} \!&=&\!
-\frac{1}{4} \, F_{\mu\nu}F^{\mu\nu}
-\frac{1}{12} \,  H_{\mu\nu\rho}H^{\mu\nu\rho}
+\frac{\theta}{2} \, \epsilon^{\mu\nu\rho\sigma} A_{\mu} \, \partial_{\nu}B_{\rho\sigma}
\nonumber \\
&&
\hspace{-0.5cm}
+\left(D_{\mu}\Phi\right)^{\dagger}
D^{\mu}\Phi
-\mu^{2} \, (\Phi^{\dagger}\Phi)
-\frac{\lambda}{6}
\left(
\Phi^{\dagger}\Phi
\right)^{2} \; ,
\end{eqnarray}
where $F_{\mu\nu}=\partial_{\mu}A_{\nu}-\partial_{\nu}A_{\mu}$ is the usual EM field strength tensor, $H_{\mu\nu\rho}=
\partial_{\mu}B_{\nu\rho}+\partial_{\nu}B_{\rho\mu}+\partial_{\rho}B_{\mu\nu}$ is the field  strength tensor for the Kalb-Ramond field that we denote by anti-symmetric tensor $B_{\mu\nu}=-B_{\nu\mu}$, and $\theta$ is a topological parameter with mass dimension that mixes the $A^{\mu}$-potential with the $2$-form $B^{\mu\nu}$. In the scalar sector, the complex scalar field $\Phi$ is coupled to the gauge sector through the covariant derivative operator :  
\begin{equation}
D_{\mu}\Phi
=
\left( \,
\partial_{\mu}
-
i\,e\,A_{\mu}-i\,g\,\tilde{H}_{\mu}
\,\right)\Phi \; ,
\end{equation}
in which $\tilde{H}_{\mu}=\epsilon^{\mu\nu\rho\sigma}H_{\nu\rho\sigma}/6$ is the dual of the Kalb-Ramond strength tensor, $g$ is the coupling constant that has length dimension, $\mu^2$ and $\lambda$ are real constants from the scalar potential.
The action associated with the Lagrangian (\ref{LKB}) 
is $U(1)$ local gauge invariant under the couple of transformations $\Phi
\longmapsto
\Phi^{\prime}(x) = e^{ie\,\alpha(x)}\,\Phi(x)$,     
$A_{\mu}
\longmapsto
A^{\prime}_{\mu}=A_{\mu}+
\partial_{\mu}\alpha$, and
$B_{\mu\nu}
\longmapsto
B_{\mu\nu}^{\prime}=B_{\mu\nu}+\partial_{\mu}\Lambda_{\nu}-
\partial_{\nu}\Lambda_{\mu},
$
in which $\Lambda_{\mu}(x)$ is an arbitrary $4$-vector, that also transforms under a gauge symmetry by $\Lambda_{\mu} \longmapsto \Lambda^{\prime}_{\mu}=\Lambda_{\mu}+\partial_{\mu}\beta$, for $\alpha(x)$ and $\beta(x)$ two arbitrary real functions. Under these transformations, $\tilde{H}^{\mu}$ is unchanged, thus the covariant derivative transforms like the complex scalar field under $U(1)$, as the usual case. The dimensional analysis in $4D$ shows all the fields of the model have mass dimension 1, while the field strength tensors have mass dimensions 2. There by, since the $g$-coupling has length dimension (inverse of mass dimension), we have an effective model that is not non-renormalizable in all orders of perburbative QFT. However, we limit ourselves to investigate the model up to one loop correction for the effective potential.  
%

%
%
%
%
%
%
%

%
The scalar potential $V(\Phi^{\dagger}\Phi)=\mu^{2}(\Phi^{\dagger}\Phi)
+\frac{\lambda}{6}\left(\Phi^{\dagger}\Phi\right)^{2}$ develops a vacuum expectation value of $v=\sqrt{-3\,\mu^2/\lambda}$ for $\mu^2<0$. In the QFT approach for the effective potential, the effective action is defined by the Legendre transformation : $\Gamma[\phi_{c}]=W[J]-\int d^{4}x \, J(x) \, \phi_{c}(x)$, in which the {\it classical field} is given by the derivative of the $W[J]$-functional in relation to classical sources $J$, {\it i.e.}, $\phi_{c}(x)=\delta W[J]/\delta J$, where $W[J]$ is the generating functional for connected Green functions. For loop calculations in the effective potential, we consider fluctuations around the classical field $\phi_{c}\equiv v$ (constant field) in which the scalar field $\Phi$ is written as     
\begin{eqnarray}\label{Phiphi1phi2}
\Phi(x)=\frac{ \phi_{c} + \phi_{1}(x)+i\,\phi_{2}(x) }{\sqrt{2}} \; ,
\end{eqnarray}
where $\phi_{i}(x)\,(i=1,2)$ are two real scalar fields. Furthermore, the quantization of the model in the gauge sector requires the gauge fixing terms 
\begin{equation} 
\mathcal{L}_{gf} = 
-\frac{1}{2\xi} \left( \partial_{\mu}A^{\mu} \right)^2
-\frac{1}{2\zeta} \left( \partial_{\mu}B^{\mu\nu} \right)^2 \; ,
\end{equation} 
are added to the Lagrangian (\ref{LKB}), in which $\xi$ and $\zeta$ denotes the gauge-fixing parameters for the Maxwell and Kalb--Ramond gauge fields, respectively.

Substituting (\ref{Phiphi1phi2}) in the scalar sector of (\ref{LKB}), we obtain : 
\begin{eqnarray}
&&
\mathcal{L}_{scalar}=\frac12(\partial_{\mu}\phi_{1})^{2}
+
\frac12(\partial_{\mu}\phi_{2})^{2}
\nonumber\\[2mm]
&&
+
\phi_{c}
\left( e\,A_{\mu}+g\,\widetilde{H}_{\mu}\right)
\partial^{\mu}\phi_{2}
\nonumber\\[2mm]
&&
+
\frac12
\phi_{c}^{2}
\left(eA_{\mu}+g\widetilde{H}_{\mu}\right)
\left(eA^{\mu}+g\widetilde{H}^{\mu}\right)
\nonumber\\[2mm]
&&
+
(eA_{\mu}+g\widetilde{H}_{\mu})
\left(
\phi_{1}\partial^{\mu}\phi_{2}
-
\phi_{2}\partial^{\mu}\phi_{1}
\right)
\nonumber\\[2mm]
&&
+
\frac12
\,(eA_{\mu}+g\widetilde{H}_{\mu})
(eA^{\mu}+g\widetilde{H}^{\mu})
\nonumber\\[2mm]
&&
\times \left(
2\phi_{c}\phi_{1}
+\phi_{1}^{2}
+\phi_{2}^{2}
\right)
-V_{eff}^{(0)}(\phi_c)
\nonumber\\[2mm]
&&
-
\left(
\mu^{2}\phi_{c}
+
\frac{\lambda}{6}\phi_{c}^{3}
\right)\phi_{1}
\nonumber\\[2mm]
&&
-
\frac12
\left(
\mu^{2}
+
\frac{\lambda}{2}\phi_{c}^{2}
\right)\phi_{1}^{2}
\nonumber\\[2mm]
&&
-
\frac12
\left(
\mu^{2}
+
\frac{\lambda}{6}\phi_{c}^{2}
\right)\phi_{2}^{2} \; ,
\end{eqnarray}
where $V_{eff}^{(0)}(\phi_c)=\frac12\,\mu^{2}\phi_{c}^{2}+\frac{\lambda}{24}\,\phi_{c}^{4}$ is the tree level effective potential. 
Some terms can be eliminated by a gauge choice, and also by the properties of $\phi_c$. The effective potential at one loop 
requires the quadratic terms of the Lagrangian. Using the properties of the Kalb-Ramond field and its field strength tensor, the quadratic 
terms are given by
\begin{eqnarray}\label{L2}
{\cal L}^{(2)} &=& -\frac{1}{4} \, F_{\mu\nu}F^{\mu\nu}+\frac{1}{2} \, e^{2}\,\phi_{c}^2\,A_{\mu}^2-\frac{1}{2\xi}\,\left(\partial_{\mu}A^{\mu} \right)^2
\nonumber \\
&&
-\frac{1}{12} \left(1+g^2\,\phi_{c}^2\right)  H_{\mu\nu\rho}H^{\mu\nu\rho}-\frac{1}{2\zeta}\,\left(\partial_{\mu}B^{\mu\nu} \right)^2
\nonumber \\
&&
+M(\phi_{c}) \, \epsilon^{\mu\nu\rho\sigma} \, A_{\mu} \, \partial_{\nu}B_{\rho\sigma}
\nonumber \\
&&
+\frac{1}{2} \,
(\partial_{\mu}\phi_{1})^2
-
\frac{1}{2} \,
\left(
\mu^{2}
+
\frac{\lambda}{2}\,\phi_{c}^{2}
\right)
\phi_{1}^{2}
\nonumber \\
&&
+
\frac{1}{2}
(\partial_{\mu}\phi_{2})^2
-
\frac{1}{2}
\left(
\mu^{2}
+
\frac{\lambda}{6}\,\phi_{c}^{2}
\right)
\phi_{2}^{2} \; ,
\end{eqnarray}
where $M(\phi_c)$ is defined by
\begin{eqnarray}
M(\phi_c)=\frac{\theta}{2}+\frac{1}{4} \, e\,g\,\phi_{c}^2 \simeq\frac{1}{4} \, e\,g\,\phi_{c}^2 \; ,
\end{eqnarray}
and we are considering the vacuum effects through $\phi_{c}$ which have more important contribution compared to the $\theta$-parameter, 
{\it i.e.}, $g\,\phi_{c}^2 \gg \theta$. Consequently, the classical field also yields topological contributions to the model due 
to expansion (\ref{Phiphi1phi2}).   
\section{Effective potential at one loop}
\label{sec3}
The quadratic action associated with (\ref{L2}) is
\begin{eqnarray}\label{S2}
S^{(2)}=\int d^{4}x^{\prime}\,d^{4} x \left[ -\frac{1}{2} \, \phi_{i}(x^{\prime}) \, {\cal A}_{ij}(x^{\prime},x;\phi_c) \, \phi_{j}(x) \right]
\nonumber \\
+ \int d^{4}x^{\prime}\,d^{4}x \, \frac{1}{2} \, \left[N(x^{\prime})\right]^{t} {\cal M}(x^{\prime},x;\phi_c) \, N(x)
\, , \hspace{0.5cm}
\end{eqnarray}
where in the sector of the scalars $\phi_{i}\,(i=1,2)$, the operators ${\cal A}_{ij}(\bar{x}^{\prime},\bar{x};\phi_c)$ are defined
in the coordinate space as
\begin{equation}\label{Aij}
{\cal A}_{ij}(x^{\prime},x;\phi_c)=\left[\phantom{\frac{1}{2}}\!\!\!\!\!-\partial_{\bar{x}^{\prime}}^{\bar{\mu}}\partial_{\bar{x}\bar{\mu}}
+m_{\phi_{i}}^2(\phi_c) \right] \delta_{ij} \, \delta^{4}(x^{\prime}-x) \; ,
\end{equation}
with $m_{\phi_{1}}^2(\phi_c)=\mu^2+\lambda\phi_c^2/2$ , and $m_{\phi_{2}}^2(\phi_c)=\mu^2+\lambda\phi_c^2/6$. In the sector of gauge fields of 
(\ref{S2}), we write $N^{t}=(A^{\mu}\;\;B^{\kappa\lambda})$ as a column vector with components $A^{\mu}$ and $B^{\kappa\lambda}$, 
and ${\cal M}$ is set by the matrix
\begin{equation}
{\cal M}(x^{\prime},x;\phi_c)=\left[
\begin{array}{cc}
X_{\mu\nu}(x^{\prime},x;\phi_c) & S_{\mu\rho\sigma}(x^{\prime},x;\phi_c) 
\\
\\
S_{\kappa\lambda\nu}(x^{\prime},x;\phi_c) & W_{\kappa\lambda,\rho\sigma}(x^{\prime},x;\phi_c) \\
\end{array}
\right] \; ,
\end{equation}
in which the sub-matrices are given by 
\begin{subequations}
\begin{eqnarray}
X_{\mu\nu}(x^{\prime},x,\phi_c) &=& \left[ \, \theta_{\mu\nu} \left(-\partial_{x^{\prime}}^{\alpha}\partial_{x\alpha}+e^2\phi_{c}^2\right)
\right.
\nonumber \\
&&
\hspace{-2.0cm}
\left.
+\,\omega_{\mu\nu}\left( -\frac{1}{\xi} \,\partial_{x^{\prime}}^{\alpha}\partial_{x\alpha}+e^2\phi_{c}^2 \right)\right]\delta^{4}(x^{\prime}-x) \; , \;\;\;
\label{X}
\\
S_{\mu\rho\sigma}(x^{\prime},x,\phi_c) &=& M(\phi_c)\,\epsilon_{\mu\rho\sigma\nu} \, \partial_{x^{\prime}}^{\nu} \delta^{4}(x^{\prime}-x) \; ,
\label{S}
\\
W_{\kappa\lambda,\rho\sigma}(x^{\prime},x,\phi_c) &=& \left[ -\left(\,1+g^2\phi_c^{2}\,\right)(\partial_{x^{\prime}}^{\alpha}\partial_{x\alpha}) (P_{b})_{\mu\nu,\kappa\lambda} 
\right.
\nonumber \\
&&
\hspace{-1.5cm}
\left.
+ \frac{1}{\zeta} \, (-\partial_{x^{\prime}}^{\alpha}\partial_{x\alpha}) \, (P_{e})_{\mu\nu,\kappa\lambda} \right]\delta^{4}(x^{\prime}-x) \; ,
\label{W}
\end{eqnarray}
\end{subequations}
where $\theta_{\mu\nu}$, $\omega_{\mu\nu}$, $(P_{b})_{\mu\nu,\kappa\lambda}$ and $(P_{e})_{\mu\nu,\kappa\lambda}$ are the projectors in the coordinate space
\begin{subequations}
\begin{eqnarray}
&&
\theta_{\mu\nu} =\eta_{\mu\nu}-\frac{\partial_{\mu} \partial_{\nu}}{\Box}
\quad , \quad
\omega_{\mu\nu}=\frac{\partial_{\mu} \partial_{\nu}}{\Box} \; ,
\\
&&
(P_{b})_{\mu\nu,\kappa\lambda} = \frac{1}{2} \left( \theta_{\mu\kappa}\,\theta_{\nu\lambda}-\theta_{\mu\lambda}\,\theta_{\nu\kappa} \right) \; ,
\\
&&
(P_{e})_{\mu\nu,\kappa\lambda} = \frac{1}{2} \left( \, \theta_{\mu\kappa}\,\omega_{\nu\lambda}-\theta_{\mu\lambda}\,\omega_{\nu\kappa}
\right.
\nonumber \\
&&
\left.
-\theta_{\nu\kappa}\,\omega_{\mu\lambda}+\theta_{\nu\lambda}\,\omega_{\mu\kappa} \, \right) \; .
\end{eqnarray}
\end{subequations}
These projectors satisfy the closed algebra.
They satisfy the relations :
\begin{subequations}
\begin{eqnarray}
(P_{b})_{\mu\nu,\kappa\lambda}\,(P_{b})^{\kappa\lambda,}_{\;\;\;\;\;\rho\sigma}=(P_{b})_{\mu\nu,\rho\sigma} \; ,
\\
(P_{e})_{\mu\nu,\kappa\lambda}\,(P_{e})^{\kappa\lambda,}_{\;\;\;\;\;\rho\sigma}=(P_{e})_{\mu\nu,\rho\sigma} \; ,
\\
(P_{b})_{\mu\nu,\kappa\lambda}\,(P_{e})^{\kappa\lambda,}_{\;\;\;\;\;\rho\sigma}=0 \; ,
\\
(P_{e})_{\mu\nu,\kappa\lambda}\,(P_{b})^{\kappa\lambda,}_{\;\;\;\;\;\rho\sigma}=0 \; ,
\\
S_{\mu\nu\alpha}\,S^{\alpha\kappa\lambda}=-2\,\Box \, (P_{b})_{\mu\nu,}^{\;\;\;\;\;\kappa\lambda} \; ,
\\
S_{\mu}^{\;\;\,\kappa\lambda}\,S_{\nu\kappa\lambda}=-2\, M(\phi_{c})^2 \, \Box \, \theta_{\mu\nu} \; ,
\\
(P_{b})_{\mu\nu,\kappa\lambda} \, S^{\kappa\lambda\rho}=S_{\mu\nu}^{\;\;\;\,\rho} \; ,
\\
S^{\kappa\alpha\beta}\,(P_{b})_{\alpha\beta,}^{\;\;\;\;\;\;\mu\nu}=S^{\kappa\mu\nu} \; ,
\\
(P_{e})_{\mu\nu,\kappa\lambda} \, S^{\kappa\lambda\rho}=0 \; ,
\\
S^{\kappa\alpha\beta}\,(P_{e})_{\alpha\beta,}^{\;\;\;\;\;\;\mu\nu}=0 \; ,
\end{eqnarray}
\end{subequations} 
that will be useful ahead.

In (\ref{S2}), we redefine the fields as $\phi_{i} \rightarrow \sqrt{\hbar} \, \phi_{i}$, $A^{\mu} \rightarrow \sqrt{\hbar} \, A^{\mu}$, and $B^{\mu\nu} \rightarrow \sqrt{\hbar} \, B^{\mu\nu}$, in which the terms proportional to $\hbar^{1/2}$, $\hbar$, and at higher order of $\hbar$ are quantum corrections beyond the one-loop. Thus, the quadratic 
terms are important because in the definition of the generator functional we have $\hbar^{-1}$ times the correspondent action.
Therefore, the approach for the effective potential at one loop contributes to the effective action by $\Gamma[\phi_{c}]=\Gamma_{0}[\phi_c]+\hbar\, \Gamma_{1}[\phi_c]$, 
where $\Gamma_{0}[\phi_c]$ is at tree level, and $\Gamma_{1}[\phi_c]$ is the effective action at one loop, with the correction in $\hbar$-order. The expression of $\Gamma_{1}[\phi_c]$ is
\begin{eqnarray}\label{Traces}
\Gamma_{1}[\phi_c] &=& - \, \Omega \, V_{1}(\phi_c)=-\frac{i}{2} \left\{ \, \mbox{Tr} \ln\left[ \, \frac{ {\cal A}(x^{\prime},x;\phi_c) }{{\cal A}(x^{\prime},x;0)} \, \right] 
\right.
\nonumber \\
&&
\left.
+\mbox{Tr} \ln\left[ \, \frac{ {\cal M}(x^{\prime},x;\phi_c) }{{\cal M}(x^{\prime},x;0)} \, \right] \, \right\} \; ,
\end{eqnarray}
in which $\int d^{4}x=\Omega$ is the $4D$ volume, 
$\mbox{Tr}$ means the trace operation in the coordinate space, that in the gauge sector includes the trace over the 
index $(\mu\nu,\kappa\lambda)$. Using the Fourier representation of the Dirac delta in (\ref{Aij}) and in (\ref{X})-(\ref{W}),
the traces (\ref{Traces}) can be written in the momentum space with $\partial_{\mu} \rightarrow i\,k_{\mu}$ and 
$\partial_{x^{\prime}}^{\alpha}\partial_{x\alpha} \rightarrow k^{2}$. All the projectors defined previously satisfy the same closed algebra.
In the sector of the ${\cal M}$-matrix, we use the identity $\mbox{Tr}(\ln {\cal M})=\ln(\det {\cal M})$, with $\det({\cal M})=\det(W)\cdot\det(X-S\,W^{-1}\,S)$, 
where $W^{-1}$ is the inverse of (\ref{W}). Therefore, using the properties of the projectors, we obtain    
%
%
%
\begin{eqnarray}
&&
X_{\mu\nu} - S_{\mu\rho\sigma} \, (W^{-1})^{\rho\sigma,\kappa\lambda} \, S_{\kappa\lambda\nu}
\nonumber \\
&&
= \theta_{\mu\nu} 
\left[ \, \Box+e^2\,\phi_{c}^2\,f(\phi_c)\, \right]+\omega_{\mu\nu}\left[ \frac{\Box}{\xi}+e^2\,\phi_{c}^2 \right] 
\; , \hspace{0.6cm}
\end{eqnarray}
where we have defined the function $f(\phi_c)=1+g^2\phi_{c}^2/[\,8(1+g^2\phi_{c}^2)\,]$. Notice that, when $g\rightarrow 0$, $f(\phi_c)=1$ and all the contribution of 
$g$-coupling disappears from the effective potential. This is the limit to recover the Coleman-Weinberg result for the effective potential in usual scalar electrodynamics. 
Thereby, the effective potential at one loop is $V_{eff}^{(1)}(\phi_c)=V_{eff}^{(0)}(\phi_c)+V_{1}(\phi_c)$, that in the momentum space is set by the integrals

\begin{eqnarray}\label{Veff1}
V_{eff}^{(1)}(\phi_c) &=& \frac{1}{2} \, \mu^2 \, \phi_c^2+\frac{\lambda}{24} \, \phi_{c}^4
\nonumber \\
&&
\hspace{-1cm}
-\frac{i}{2} \int
\frac{d^{4}k}{(2\pi)^{4}}
\left\{ \,
3\,
\ln\left[ \, 
\frac{-k^{2}+e^{2}\,\phi_{c}^{2} \, f(\phi_c) }{-k^2+\theta^2}
\,\right]
\right.
\nonumber \\
&&
\hspace{-1cm}
\left.
+
\ln\left[\, \frac{-k^2+\mu^2
\lambda \,\phi_{c}^{2}/2}{-k^2+\mu^2}
\, \right]
\right.
\nonumber \\
&&
\hspace{-1cm}
\left.
+
\ln\left[\, \frac{-k^2+\mu^2+\lambda \,\phi_{c}^{2}/6}{-k^2+\mu^2}
\,\right]
\right.
\nonumber \\
&&
\hspace{-1cm}
\left.
+3\,\ln\left( 1+g^2\,\phi_{c}^2 \right) \phantom{\frac{1}{2}} \!\!\!\!
\right\} \, . \;\;\;    
\end{eqnarray}
These integrals are ultraviolet divergent (and also infrared divergent in the case of the last line) in $4$-dimensions. In the next section,  
we will use the dimensional regularization $(D)$ to make it finite, and after that, the physical result is recovered when $D \rightarrow 4$.

\section{Regularization of the effective potential}
\label{sec4}

In $D$-dimensions, all the integrals of (\ref{Veff1}) are finite, but this not is the physical case. In fact, we need to modify the integrals to isolate 
the divergent terms, and then, using a renormalization scheme, we can remove these spurious terms. Furthermore, we also use conveniently the integrals in the Euclidian 
space, with $x^{0}=-i\,x_{4}$, and in the momentum space is equivalent to $k^{2}=-k_{E}^2$, where $k_{E}^2=k_{4}^2+k_{1}^2+k_{2}^2+k_{3}^2>0$, and $d^{D}k=-i\,d^{D}k_{E}$.
The modified effective potential in $D$-dimensions is 
\begin{eqnarray}
&&
V_{eff}^{(1)}(\phi_c,D) = \frac{1}{2} \, \mu^2 \, \phi_c^2+\frac{\lambda}{24} \, \phi_{c}^4
\nonumber \\
&&
-\frac{1}{2} \int
\frac{d^{D}k_{E}}{(2\pi)^{D}}
\left\{ \,
3\,
\ln\left[ \,
\frac{k_{E}^{2}+e^{2}\, \Lambda^{2-D/2}\,\phi_{c}^{2} \, f(\phi_{c}) }{k_{E}^2+\theta^2}
\, \right]
\right.
\nonumber \\
&&
\left.
+
\ln\left[1+
\frac{\lambda \, \Lambda^{2-D/2} \, \phi_{c}^{2}/2}{k_{E}^2+\mu^2}
\right]
\right.
\nonumber \\
&&
\left.
+
\ln\left[1+
\frac{\lambda \, \Lambda^{2-D/2} \,\phi_{c}^{2}/6}{k_{E}^2+\mu^2}
\right]
+3\,\ln(1+g^2\,\phi_{c}^2) \,
\right\} \, , \;\;\;\; \;\;\;\;   
\end{eqnarray}
where $\Lambda$ is an arbitrary energy scale, and the coupling constants $e$ and $\lambda$ remain dimensionless in $D$-dimensions. Using the technicalities of dimensional regularization, the result of the integrals is given by
\begin{eqnarray}\label{Veff1resultD}
&&
V_{eff}^{(1)}(\phi_c,D)= \frac{1}{2} \, \mu^2 \, \phi_c^2+\frac{\lambda}{24} \, \phi_{c}^4
\nonumber \\
&&
-\frac{1}{2\,(4\pi)^{D/2}} \, \Gamma\left(-\frac{D}{2}\right)
\nonumber \\
&&
\times \left\{ \, 3 \, (\theta^{2})^{D/2}
-3\left[\, e^{2}\, \Lambda^{2-D/2}\,\phi_{c}^{2}\,f(\phi_c) \, \right]^{D/2}
\right.
\nonumber \\
&&
\left.
+2\,(\mu^2)^{D/2}
-\left[\,\mu^2+\frac{\lambda}{2} \, \Lambda^{2-D/2} \, \phi_c^2 \, \right]^{D/2}
\right.
\nonumber \\
&&
\left.
-\left[\,\mu^2+\frac{\lambda}{6} \, \Lambda^{2-D/2} \, \phi_c^2 \, \right]^{D/2}
\, \right\} \; ,
\end{eqnarray}
that is valid in all complex plane of $D$, except for $\Re[D]=\left\{ \, 0 \, , \, 2 \, , \, 4 \, , \, 6 \, \ldots  \right\}$, in accord with the properties of the Gamma function. The physical dimension is so recovered when $D=4-\epsilon$, expanding the result (\ref{Veff1resultD}) around the $\epsilon \rightarrow 0$. For a small $\mu^2$- and $\theta^2$-parameters in which $\lambda\,\phi_{c}^2 \gg \left( \,\mu^2 \, , \, \theta^2 \,\right)$, the result is :
\begin{eqnarray}\label{Veff1renor}
&&
V_{eff}^{(1)}(\phi_c,\epsilon)=\frac{1}{2}\,\mu^2\,\phi_c^2+\frac{\lambda}{24} \, \phi_{c}^4
\nonumber \\
&&
-\frac{3 \, e^{4} \, \phi_{c}^4}{32 \,\pi^2} \, f(\phi_c)^2 \,\frac{1}{\epsilon} -\frac{5 \lambda^2 \, \phi_{c}^4}{576 \,\pi^2}\frac{1}{\epsilon}
\nonumber \\
&&
-\left[ \, \frac{3}{4}-\frac{\gamma}{2}+\ln(2\sqrt{\pi}) \, \right] \left[ \, \frac{3 \, e^4 \, \phi_c^4}{32 \pi^2} \, f(\phi_c)^2+\frac{5 \lambda^2 \, \phi_{c}^4}{576 \pi^2}  \, \right]
\nonumber \\
&&
+\frac{\lambda^2 \, \phi_{c}^4}{2304\pi^2} \, \ln\left( \frac{\lambda \,\phi_{c}^2}{6\Lambda^2} \right)
+\frac{\lambda^2 \, \phi_{c}^4}{256\pi^2} \, \ln\left( \frac{\lambda \,\phi_{c}^2}{4\Lambda^2} \right)
\nonumber \\
&&
+\frac{3\,e^4\,\phi_{c}^4}{64\pi^2} \, f(\phi_c)^2 \, \ln\left[\,\frac{e^2\,\phi_c^2}{\Lambda^2}\,f(\phi_c)\,\right] \; ,
\end{eqnarray}
where $\gamma=0.577$ is the Euler-Mascheroni constant. Taking the approximation of $g^2\,\phi^2_c \ll 1$, the effective potential in this regime is
\begin{eqnarray}\label{Veff1renorsimp}
&&
V_{\mathrm{eff}}^{(1)}(\phi_{c},\epsilon)
=\frac{1}{2}\,\mu^2\,\phi_c^2+\frac{\lambda}{24}\phi_{c}^{4}
-\frac{3e^{4}\phi_{c}^{4}}{32\pi^{2}}\frac{1}{\epsilon}-
\frac{5\lambda^{2}\phi_{c}^{4}}{576\pi^{2}}\frac{1}{\epsilon}
\nonumber \\
&&
-\left[
\frac{3}{4}
-\frac{\gamma}{2}
+
\ln\left(2\sqrt{\pi}\right)
\right]
\Bigg(
\frac{3e^{4}\phi_{c}^{4}}{32\pi^{2}}
+
\frac{5\lambda^{2}\phi_{c}^{4}}{576\pi^{2}}
\Bigg)
\nonumber \\
&&
+
\frac{\lambda^{2}\phi_{c}^{4}}{2304\pi^{2}}
\ln\left(
\frac{\lambda\phi_{c}^{2}}{6\Lambda^{2}}
\right)
+
\frac{\lambda^{2}\phi_{c}^{4}}{256\pi^{2}}
\ln\left(
\frac{\lambda\phi_{c}^{2}}{4\Lambda^{2}}
\right)
\nonumber \\
&&
+
\frac{3e^{4}\phi_{c}^{4}}{64\pi^{2}}
\ln\left(
\frac{e^{2}\phi_{c}^{2}}{\Lambda^{2}}
\right)
-g^2 \, \phi_c^6 \, \Bigg\{ \, \frac{3e^{4}}{128\pi^{2}}\frac{1}{\epsilon}
\nonumber \\
&&
-\frac{3e^{4}}{128\pi^{2}}
\left[
\frac{3}{4}
-\frac{\gamma}{2}
+\ln(2\sqrt{\pi})
\right]
\nonumber\\
&&
+\frac{3e^{4}}{256\pi^{2}}
\ln\left(
\frac{e^{2}\phi_c^{2}}{\Lambda^{2}}
\right)
+\frac{3e^{4}}{512\pi^{2}} \, \Bigg\} \; .
\end{eqnarray}
The results (\ref{Veff1renor}) and (\ref{Veff1renorsimp}) show the divergent terms in $1/\epsilon$, when $\epsilon=0$. These spurious  terms will be removed through minimal renormalization scheme in the next section.   
\begin{figure}[t]
    \centering
    \includegraphics[width=1.05\linewidth]{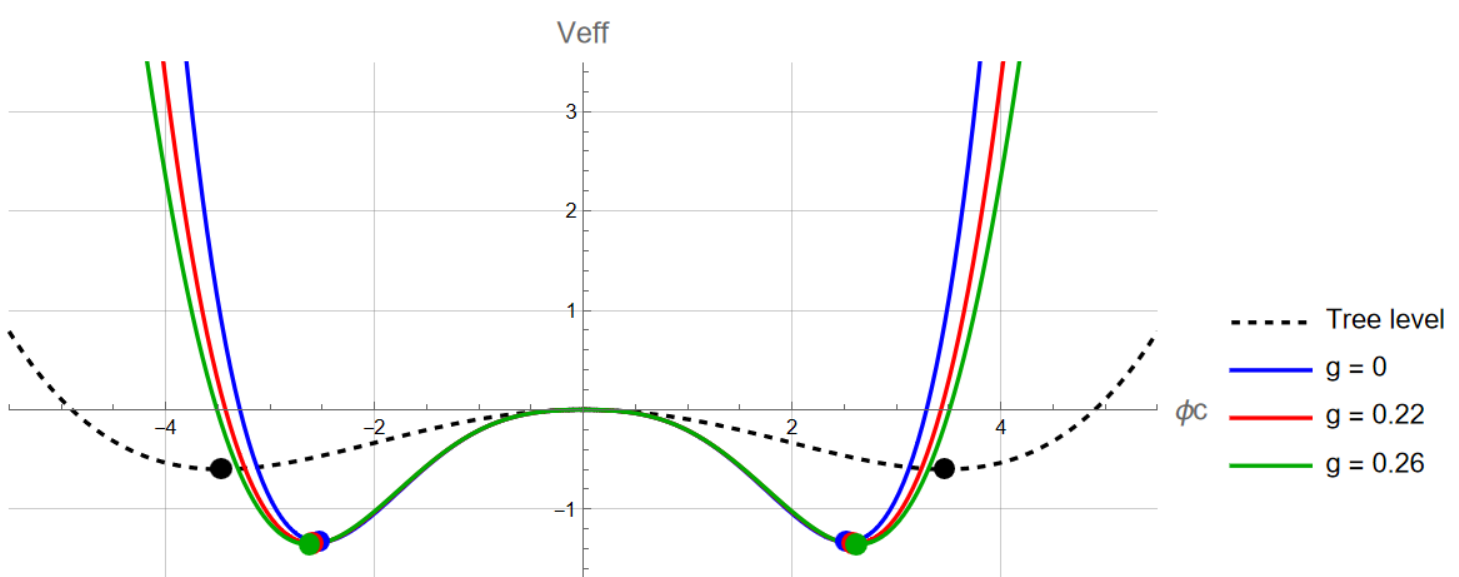}
    \caption{Tree-level and one-loop renormalized effective potential $V_{\mathrm{eff}}^{ren}(\phi_c)$ as functions of the classical scalar field $\phi_c$ for some values of the $g$-coupling. The one-loop corrections modify the shape of the classical potential while preserving the qualitative features of the vacuum structure. We use the values of $\mu^2=-0.2$ , $\lambda=0.1$, $e=1.8$, $\Lambda=1$ (in energy dimensions) for this plot.}
    \label{fig:veff}
\end{figure}
%
\section{Renormalization of the effective potential at one loop}
\label{sec5}

We now proceed with the renormalization of the effective potential at one loop. Our considerations for the renormalization are minimal, in the sense that we only remove the divergent part of the effective potential by introducing counter terms, while the finite parts are left untouched. Thus, we apply the MS renormalization scheme where the subtraction is done at $\phi_c=0$, instead of a physical renormalization of the full potential. Note, the Lagrangian already has a negative dimension coupling $g$. This essentially means that the theory is non-renormalizable and can be at most be treated as an effective one. Indeed, the divergent part of the one loop effective potential has a term proportional to $\phi^6$, whereas such a term is not there in the tree level potential. To remove this divergence, we have to introduce a new counter term and add it to the tree level action. 
The divergent part of the one-loop effective potential is 
\begin{equation} 
V_{\mathrm{div}}^{(1)}(\phi_c) = -\frac{1}{\epsilon} \left( \frac{3e^4}{32\pi^2} + \frac{5\lambda^2}{576\pi^2} \right)\phi_c^4 - \frac{3e^4g^2 \, \phi_c^6}{128\pi^2} \frac{1}{\epsilon} \; . 
\end{equation} 
The counter-term potential is taken to be 
\begin{equation} 
V_{\mathrm{ct}}(\phi_c) = \frac{\delta\lambda}{4!}\phi_c^4 + \frac{\delta\kappa}{6!}\phi_c^6 \; , 
\end{equation} 
where $\delta\lambda$ and $\delta\kappa$ are shifts from the original couplings which are needed to absorb the divergent parts of the effective potential. We define the pole part of the effective potential by 
\begin{equation} 
V_{\mathrm{pole}}(\phi_c) = V_{\mathrm{div}}^{(1)}(\phi_c) + V_{\mathrm{ct}}(\phi_c) \; . 
\end{equation} 
In the pure minimal subtraction scheme, the sextic pole is removed by imposing the renormalization condition 
\begin{equation} 
\frac{d^6V_{\mathrm{pole}}}{d\phi_c^6}\Bigg|_{\phi_c=0} = 0 \; . 
\end{equation} 
It yields the counter-term 
%
%
%
\begin{equation} 
\delta\kappa = \frac{135e^4g^2}{8\pi^2} \frac{1}{\epsilon} \; . 
\end{equation} 
The quartic pole is removed by imposing the next renormalization condition
\begin{equation} \frac{d^4V_{\mathrm{pole}}}{d\phi_c^4}\Bigg|_{\phi_c=0} = 0 \; . 
\end{equation} 
The sextic contribution cancels in this combination, and the condition becomes 
\begin{equation} 
-\frac{24}{\epsilon} \left( \frac{3e^4}{32\pi^2} + \frac{5\lambda^2}{576\pi^2} \right) + \delta\lambda = 0 \; . \end{equation} 
Therefore, 
\begin{equation} 
\delta\lambda = \frac{9e^4}{4\pi^2} \frac{1}{\epsilon} + \frac{5\lambda^2}{24\pi^2}\frac{1}{\epsilon} \; . 
\end{equation}
Since there is no $\mu^2$-dependency on the one loop part of the effective potential, there is no divergence at quadratic order in $\phi_c$. Thus, essentially, we do not need to renormalize the parameter $\mu^2$. We next define 
\begin{equation} 
K = \frac{3}{4} - \frac{\gamma}{2} + \ln\left(2\sqrt{\pi}\right) \simeq 1.72 \; . 
\end{equation} 
After the subtraction of the poles, the renormalized effective potential in the pure minimal subtraction scheme is 
\begin{eqnarray}\label{Veff1finite}
&&
V_{\mathrm{eff}}^{\mathrm{ren}}(\phi_c) = \frac{1}{2}\,\mu^2\,\phi_c^2+ \frac{\lambda}{24}\phi_c^4 
\nonumber \\ 
&&
- K \left[ \frac{3e^4\phi_c^4}{32\pi^2} \left( 1-\frac{g^2\phi_c^2}{4} \right) + \frac{5\lambda^2\phi_c^4}{576\pi^2} \right] 
\nonumber \\ 
&&
+ \frac{\lambda^2\phi_c^4}{2304\pi^2} \ln\left( \frac{\lambda\phi_c^2}{6\Lambda^2} \right) 
+ \frac{\lambda^2\phi_c^4}{256\pi^2} \ln\left( \frac{\lambda\phi_c^2}{4\Lambda^2} \right) 
\nonumber \\ 
&&
+ \frac{3e^4 \, \phi_c^4}{64\pi^2} \left[ \, \ln\left( \frac{e^2\phi_c^2}{\Lambda^2} \right) - \frac{g^2\phi_c^2}{4} \ln\left( \frac{e^2\phi_c^2}{\Lambda^2} \right) - \frac{g^2\phi_c^2}{8} \, \right] \; .
\nonumber \\ 
\end{eqnarray} 
\\
The one-loop effective potential develops a divergence proportional to $\phi_c^6$, which is removed by introducing the counter-term $\delta\kappa\,\phi_c^6/6!$. In the minimal-subtraction scheme, we do not retain an independent finite tree-level $\phi_c^6$-interaction and impose the renormalization condition ($\kappa(\mu_0)=0$) at the reference scale ($\mu_0$). We identify this reference scale with the energy scale $\Lambda$. Consequently, after cancellation of the ($1/\epsilon$) pole, the renormalized effective potential consists of the original tree-level potential together with the finite one-loop contributions, including the finite terms proportional to $g^2\,\phi_c^6$. The tree level (dashed line) and the renormalized effective potential (\ref{Veff1finite}) are plotted as functions of the classical field $(\phi_{c})$ in the figure (\ref{fig:veff}), for the values of $g=0$ (blue line), $g=0.22$ (red line) and $g=0.26$ (green line). The figure shows that the increase of the $g$-coupling changes the minimal for the potential, that consequently, increases the VEV scale.    
\section{conclusions}
\label{sec6}
We investigate the properties of the effective potential at one loop associated with the Kalb-Ramond-Maxwell electrodynamics coupled to a complex scalar field. Kalb-Ramond is an antisymmetric gauge field (2-form) that arises from massless excitations of closed strings. This exotic gauge field defines a field strength tensor with three space-time index totally anti-symmetric that is gauge invariant, and its dual is a $4$-vector in four dimensions. We introduce the coupling of a complex scalar field with the electromagnetic and dual $4$-vector of Kalb-Ramond to study the effective potential of the model in one loop through the functional formalism of field theory. Since this coupling with Kalb-Ramond has a length dimension, we have an effective model in which the renormalization can be performed at one loop approximation.  
We expand the complex scalar for fluctuations around the classical field $(\phi_c)$, where the effective potential at one loop is obtained by the functional integration of the quadratic terms for the resultant Lagrangian density. Since it is known from QFT, the effective potential is divergent, we introduced dimensional regularization in the integrals. After taking the limit for physical dimension, we isolate the divergent terms, and using a minimal renormalization scheme, we obtain the finite effective potential at one loop approximation. For weak coupling with Kalb-Ramond, the result shows that the effective potential depends on $\phi_{c}^{6}$, beyond the usual quartic coupling $\phi_{c}^4$. This new behaviour is due to the topological mixing of the electromagnetic with the Kalb-Ramond field. Thereby, we illustrate the effective potential as function of the classical field for some values of the Kalb-Ramond coupling in the figure (\ref{fig:veff}). The potential curves keep the structure 
of the minimal points associated with the stable vacuum expectation value that increases with the Kalb-Ramond coupling. As perspective for a forthcoming project, the study of the renormalization group, such as the beta function and anomalous dimension using a physical renormalization scheme can be a new investigation source for the running coupling constants with an arbitrary energy scale.

\end{document}